\documentclass[aps,showpacs,amssymb,twocolumn,prl,superscriptaddress]{revtex4-1}
\usepackage{amsmath}
\usepackage{mathtools}
\usepackage{hyperref}
\usepackage[dvipdfmx]{graphicx}
\usepackage{bbm}
\usepackage{color}

\usepackage{lineno}

\usepackage{bm}

\allowdisplaybreaks

\begin{document}

\title{Generation of current vortex by spin current in Rashba systems}

\author{Florian Lange}
\email[]{langef@uni-greifswald.de}
\affiliation{Institut f\"ur Physik, Universit\"at
Greifswald, 17489 Greifswald, Germany}

\author{Satoshi Ejima}
\affiliation{Institut f\"ur Physik, Universit\"at
Greifswald, 17489 Greifswald, Germany}
\affiliation{
%Computational Condensed Matter Physics Laboratory, 
RIKEN Cluster for Pioneering Research (CPR), Wako, Saitama 351-0198, Japan}

\author{Junji Fujimoto}
\affiliation{Kavli Institute for Theoretical  Sciences, University of Chinese Academy of Sciences, Beijing, 100190, China}

\author{Tomonori Shirakawa}
\affiliation{
%Computational Materials Science Research Team, 
RIKEN Center for Computational Science (R-CCS), 
Kobe, Hyogo 650-0047, Japan}

\author{\\Holger Fehske}
\affiliation{Institut f\"ur Physik, Universit\"at
Greifswald, 17489 Greifswald, Germany}

\author{Seiji Yunoki}
\affiliation{
%Computational Materials Science Research Team, 
RIKEN Center for Computational Science (R-CCS), 
Kobe, Hyogo 650-0047, Japan}
\affiliation{
%Computational Condensed Matter Physics Laboratory, 
RIKEN Cluster for Pioneering Research (CPR), Wako, Saitama 351-0198, Japan}
\affiliation{
%Computational Quantum Matter Research Team, 
RIKEN Center for Emergent Matter Science (CEMS), Wako, Saitama 351-0198, Japan}

\author{Sadamichi Maekawa}
\email[]{sadamichi.maekawa@riken.jp}
\affiliation{
%Computational Quantum Matter Research Team, 
RIKEN Center for Emergent Matter Science (CEMS), Wako, Saitama 351-0198, Japan}
\affiliation{Kavli Institute for Theoretical  Sciences, University of Chinese Academy of Sciences, Beijing, 100190, China}

\begin{abstract}
  Employing unbiased large-scale time-dependent density-matrix renormalization-group simulations, we demonstrate the generation of a charge-current vortex via spin injection in the Rashba system.    
  The spin current is polarized perpendicular to the system plane and injected from an attached antiferromagnetic spin chain. We discuss the conversion between spin and orbital angular momentum in the current vortex that occurs because of the conservation of the total angular momentum and the spin-orbit interaction.  This is in contrast to the spin Hall effect, in which the angular-momentum conservation is violated. 
   Finally, we predict the electromagnetic field that accompanies the vortex with regard to possible future experiments. 
\end{abstract}

\date{\today}

\maketitle

The interconversion of charge and spin degrees of freedom is a key issue in spintronics~\cite{maekawa2015spin}. 
Noteworthy phenomena in this regard are the spin Hall effect, which describes the generation of a transverse spin current by a charge current, and its inverse~\cite{Hirsch99,Murakami2003,IntrinsicSHE,RevModPhys.87.1213}. These effects are due to a spin asymmetry of conduction electrons by the spin-orbit coupling. A typical model for studying the spin-charge interconversion is the two-dimensional electron gas with Rashba spin-orbit coupling~\cite{Rashba84,RashbaPerspectives}.  Various effects due to the Rashba spin-orbit coupling have been extensively investigated, including the spin-orbit torque~\cite{RevModPhys.91.035004} and the Edelstein effect~\cite{Edelstein,MicroscopicEdelstein}. 
While the spin Hall conductivity actually vanishes in the Rashba model with quadratic dispersion~\cite{PhysRevB.70.041303,PhysRevB.71.245318,PhysRevB.70.201309,PhysRevB.73.195307},
spin Hall physics may still be observed in mesoscopic Rashba systems. 
It was shown, for example, that a charge current in a nanowire can induce spin accumulation at the lateral edges~\cite{PhysRevLett.95.046601}.

  In this Letter, we investigate a junction in which a spin current is transmitted into a Rashba system from an antiferromagnetic spin-1/2 Heisenberg chain. 
The spin current in the spin chain is carried by elementary excitations called spinons~\cite{QuantumMagnetism}. 
Our goal is to demonstrate the conversion of this spinon spin current into a conduction-electron spin current in the Rashba system and, in particular, to investigate the charge-current signal caused by the interplay of the spin injection and spin-orbit coupling. 
Although the junction is an interacting quantum system, it can nevertheless be efficiently simulated by using matrix-product-state methods~\cite{White92,PhysRevLett.91.147902,ReviewSchmitteckert} combined with a Lanczos transformation of the Rashba system~\cite{BLDMRG,BlockLanczosImpurity,LanczosImpurityReview,BLDMRGRashba}, allowing us to obtain unbiased numerical results for the current dynamics. Most notably, we show that when a spin current with spin polarization perpendicular to the system is injected at a point-like contact into the Rashba system, a charge-current vortex emerges. This is similar to the spin-charge conversion in the inverse spin Hall effect. 
What is different in our model, however, is that the direction of the current is not uniform and the system instead has a rotational symmetry around the injection point. 
The junction thus has a conserved total angular momentum, and it turns out that the injected spin angular momentum is mostly converted to orbital angular momentum of the current vortex. 
 We focus on a model with an antiferromagnetic spin chain as a spin injector. As discussed in the Supplemental Material~\cite{SuppMat}, the generation of the charge-current vortex could also be observed in other settings. 
At the end, we will discuss the relevance of our results for possible experiments.

\begin{figure}[t]
  \centering
  \includegraphics[width=0.99\linewidth]{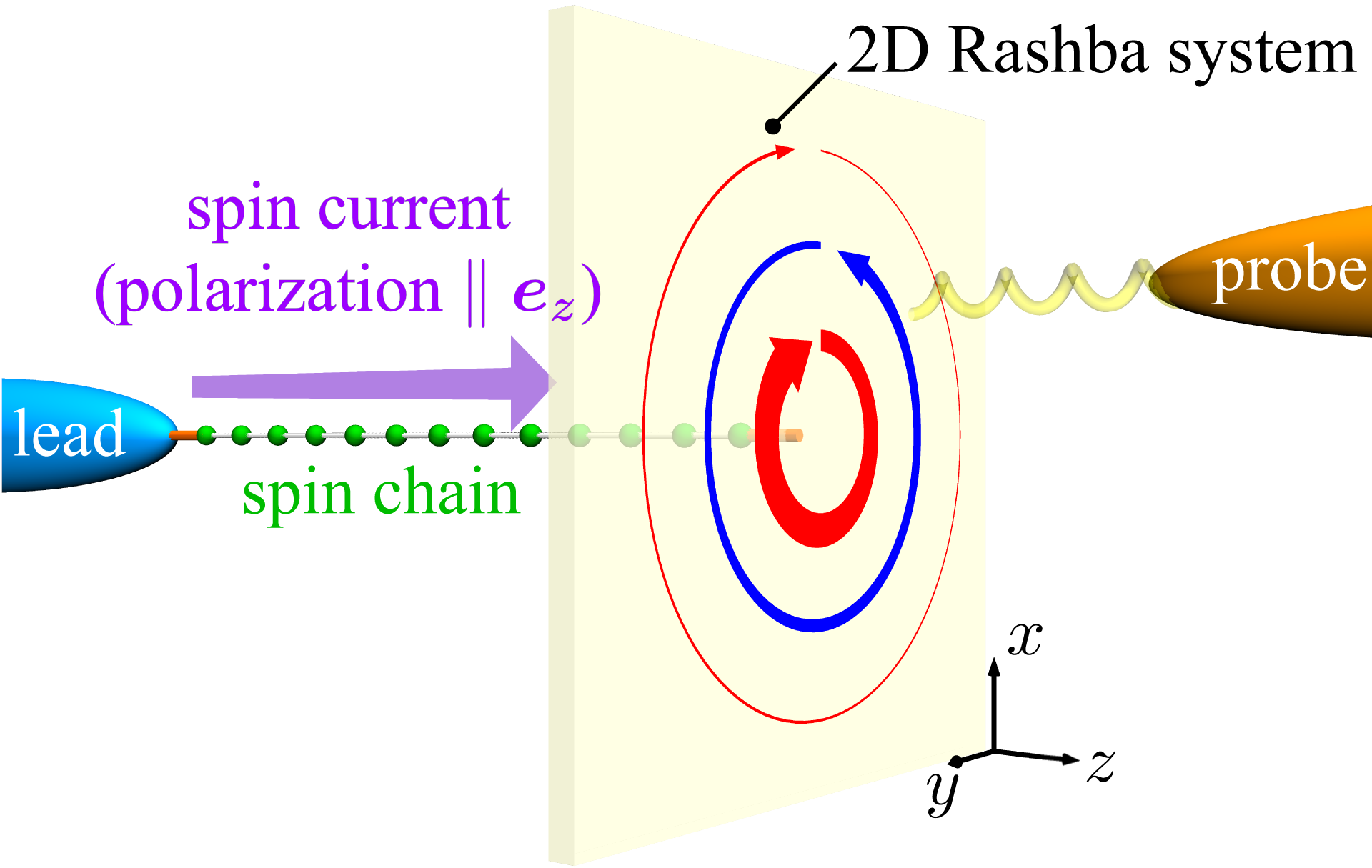}
  \caption{Sketch of the setup described by Eqs.~\eqref{eqRashbaLattice}-\eqref{eqCoupling}. 
 A spin current (purple arrow) polarized perpendicular to the Rashba plane is induced in the spin chain by switching on a spin voltage in the lead. This spin current is injected into the Rashba system, where it causes the formation of a charge-current vortex (red and blue arrows). The orange segments denote the coupling between the spin chain and the lead and Rashba systems. In an experiment, the magnetic field induced by the charge current may be detected using scanning probe microscopy.
  }
  \label{figsys}
\end{figure}

Let us first introduce the setup in more detail.  
We consider a Rashba model in the $xy$-plane on an infinite square lattice with sites $\bm{r} \in \mathbb{Z}^2$: 
  \begin{align}
  \hat{H}_R = &-\mu \sum_{\bm{r}} \sum_{\sigma = \uparrow, \downarrow} \hat{c}_{\bm{r},\sigma}^\dagger \hat{c}_{\bm{r},\sigma}^{\phantom{\dagger}} -t_R \sum_{\langle \bm{r} \bm{r}' \rangle} \sum_{\sigma = \uparrow, \downarrow} \hat{c}_{\bm{r},\sigma}^\dagger \hat{c}_{\bm{r}',\sigma}^{\phantom{\dagger}}  \nonumber \\ & - i \lambda \sum_{\bm{r}} \left( \bm{\hat{c}}_{\bm{r}}^{\dagger} \sigma^{y} \bm{\hat{c}}_{\bm{r} + \bm{e}_x}^{\phantom{\dagger}} - \bm{\hat{c}}_{\bm{r}}^{\dagger} \sigma^{x} \bm{\hat{c}}_{\bm{r} + \bm{e}_y} - \text{H.c.} \right) \, ,
  \label{eqRashbaLattice}
  \end{align}
where $\mu$ is the chemical potential, $t_R$ is the hopping, $\lambda$ is the spin-orbit-coupling strength, $\sigma^{x}$ and $\sigma^{y}$ are Pauli matrices, and $\hat{c}_{\bm{r}} = (\hat{c}_{\bm{r} \uparrow}, \hat{c}_{\bm{r}\downarrow})^{\text{T}}$ are fermion annihilation operators. 
One site $\bm{r}_0$ shall be coupled to another system that is used to inject a spin current polarized in the $z$-direction (see Fig.~\ref{figsys}). Specifically, we employ an antiferromagnetic spin-1/2 Heisenberg chain of length $N_S$,
\begin{align}
  \hat{H}_S &= J \sum_{j\geq 1}^{N_S-1} \bm{\hat{S}}_j \hat{\bm{S}}_{j+1} \, , \quad J > 0 \, .
  \label{eqSpinChain}
\end{align}
To generate a spin-current flow, the other end of the spin chain is connected to a one-dimensional semi-infinite tight-binding chain that serves as a spin reservoir:
\begin{align}
  \hat{H}_L(t) = &-t_L \sum_{j \geq 1} \sum_{\sigma} \left( \hat{c}_{j, \sigma}^\dagger \hat{c}_{j+1, \sigma}^{\phantom{\dagger}} + \text{H.c.} \right) \nonumber \\ & - \Theta(t) \frac{V}{2} \sum_{j \geq 1} \left( \hat{c}_{j, \uparrow}^\dagger \hat{c}_{j, \uparrow}^{\phantom{\dagger}} - \hat{c}_{j, \downarrow}^\dagger \hat{c}_{j, \downarrow}^{\phantom{\dagger}} \right) \, .
  \label{eq1DLead}
\end{align}
The second term in Eq.~\eqref{eq1DLead} describes a spin voltage that is switched on at time $t = 0$. 
Finally, the coupling between the subsystems is given by 
\begin{align}
  \hat{H}_C &= \frac{J'}{2} \ \sum_{\mathclap{\nu = x,y,z}} \hat{S}_{N_S}^\nu \, ( \bm{\hat{c}}_{\bm{r}_0}^\dagger \sigma^\nu \bm{\hat{c}}_{\bm{r}_0}^{\phantom{\dagger}} ) +  \frac{J''}{2} \ \sum_{\mathclap{\nu = x,y,z}} \hat{S}_{1}^\nu \, ( \bm{\hat{c}}_{1}^\dagger \sigma^\nu \bm{\hat{c}}_{1}^{\phantom{\dagger}} )  \, 
  \label{eqCoupling}
\end{align}
with $J',J'' >0$, i.e., an antiferromagnetic Heisenberg interaction. The complete Hamiltonian then becomes $\hat{H}(t) = \hat{H}_R + \hat{H}_S + \hat{H}_L(t) + \hat{H}_C$. 
It is assumed that the composite system is initially in the ground state of $\hat{H}(t < 0)$ until the spin voltage is switched on. 
Throughout this paper, we use $t_R$ as the unit of energy and set $N_S = 12$, $J = t_L = 2$, $\mu = -3.5$, and $V = 0.5$. 
Since $\hat{H}(t)$ conserves the particle number in each tight-binding system, no charge current is injected in addition to the spin current. 
We are interested in the charge current that instead develops as a consequence of the injected spin current and the spin-orbit coupling. 
Here, the charge-current-density operators for neighboring sites $\bm{r}$ and $\bm{r}+\bm{e}_{x,y}$ are defined by
 $\hat{j}_{\bm{r},\bm{r}+\bm{e}_x}^c = \bm{\hat{c}}_{\bm{r}}^\dagger (-it_R I + \lambda \sigma^y ) \bm{\hat{c}}_{\bm{r}+\bm{e}_x}^{\phantom{\dagger}} + \text{H.c.}$ and 
$\hat{j}_{\bm{r},\bm{r}+\bm{e}_y}^c = \bm{\hat{c}}_{\bm{r}}^\dagger (-it_R I - \lambda \sigma^x ) \bm{\hat{c}}_{\bm{r}+\bm{e}_y}^{\phantom{\dagger}} + \text{H.c.}$, with $I$ being the unit matrix in spin space, 
so that the total current at site $\bm{r}$ is
\begin{align}
  \bm{\hat{j}}_{\bm{r}}^c = \frac{1}{2} \left[( \hat{j}_{\bm{r},\bm{r}+\bm{e}_x}^c + \hat{j}_{\bm{r}-\bm{e}_x,\bm{r}}^c) \bm{e}_x +  (\hat{j}_{\bm{r},\bm{r}+\bm{e}_y}^c + \hat{j}_{\bm{r}-\bm{e}_y,\bm{r}}^c ) \bm{e}_y \right] \, .
 \end{align}

\begin{figure}[t]
  \centering
  \includegraphics[width=0.95\linewidth]{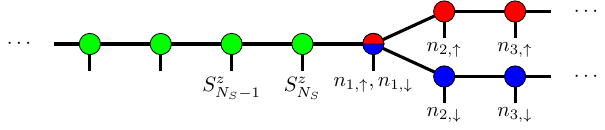}
  \caption{Tensor-network-state ansatz for the numerical simulations. The vertical lines denote the physical indices, i.e., the basis states of the local Hilbert spaces. Here, they correspond to the occupation numbers $n_{j,\sigma}$ of the fermions in the Lanczos basis or, in the spin chain, the $z$ components $S_j^z$ of the spins. The remaining lines indicate the bond indices of the tensor network. On the left side, the one-dimensional lead is similarly split into two branches (not shown). }
  \label{figTN}
\end{figure}

In order to simulate the above model numerically, we use a Lanczos transformation that maps the two-dimensional Rashba system to a chain representation~\cite{BLDMRG,BLDMRGRashba}. 
The Hamiltonian then becomes purely one-dimensional and matrix-product-state techniques can be used to calculate the ground state and simulate the time evolution with high accuracy~\cite{White92,PhysRevLett.91.147902,ReviewSchmitteckert}. To be precise, we utilize a tensor-network representation in which each tight-binding chain is split into two branches corresponding to different spin indices (pseudospin indices for the Rashba case)~\cite{TNImpurity,BLDMRGRashba}. This significantly reduces the numerical effort compared with a regular matrix-product state. 
Figure~\ref{figTN} displays the tensor network in the usual graphical notation.

For the numerical calculations, the tight-binding chain and the Lanczos representation of the Rashba system are each truncated to 500 sites. The time evolution is carried out using the time-evolving block decimation with a second-order Suzuki-Trotter decomposition and a time step $0.025$~\cite{PhysRevLett.91.147902}. For all simulated times the truncation error is kept below $10^{-7}$. In the Supplemental Material.~\cite{SuppMat}, the Lanczos transformation and the accuracy of the numerical results are discussed in further detail. 

\begin{figure}[t]
  \centering
  \includegraphics[scale=0.49]{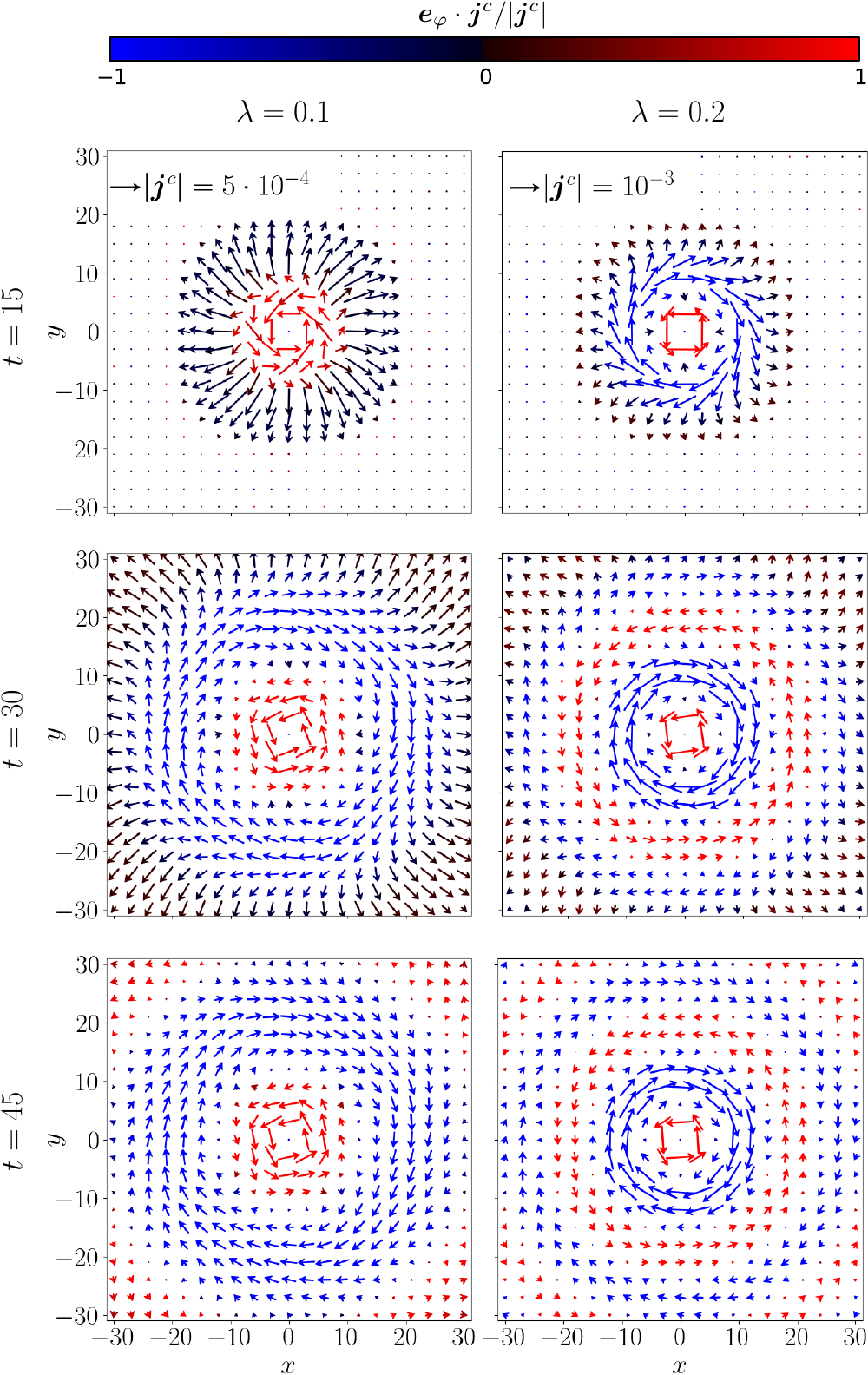} 
  \caption{Snapshot of the charge-current densities $\bm{j}_{\bm{r}}^c$ at different times $t$. For easier visualization, each arrow corresponds to the average value of the currents in a square of $3\times 3$ sites. 
    The length and color of the arrows indicate the magnitude and direction of the current, respectively. 
    Black arrows show that the current points in the radial direction, while blue (red) arrows denote current in the clockwise (counterclockwise) azimuthal direction. 
  }
  \label{figCurrent}
\end{figure}

When the spin voltage is switched on in the first lead, a spin current starts to flow at the interface with the spin chain. The perturbation spreads through the chain, approximately with the spinon velocity $J \pi / 2$, and finally reaches the Rashba system. 
At low temperatures, the efficiency of the spin injection into the Rashba system depends strongly on the coupling $J'$~\cite{ConductingFixedPoint3,ContactedSpinChain,Lange_2019}.
We have chosen $J'/J=2.15$ and $J''/J=1.70$ in order to maximize the spin current in the steady state. 
For these parameters, the spin current into the Rashba system quickly saturates 
to a value slightly below $V/(4 \pi)$, which is the current corresponding to the expected linear spin conductance of the junction with ideal contacts. 
In the following, we analyze the charge current induced by this continuous spin-current injection. We assume that the spin current is polarized in the $z$-direction. Results for an $x$-polarized spin current are presented and briefly discussed in the Supplemental Material.~\cite{SuppMat}. 

Figure~\ref{figCurrent} shows the numerically calculated charge-current profile for spin-orbit coupling parameters $\lambda = 0.1$ and $0.2$, and different simulated times $t$.
Clearly, multiple rings with circular charge current develop and then persist for long times.
Neighboring rings have opposite orientation; i.e., the current alternates between clockwise and counterclockwise.  
This behavior can be understood qualitatively as follows: 
A spin current in the Rashba system generates a transverse charge current via the inverse spin Hall effect~\cite{InverseSHE}. Here, the spin current points in the radial direction relative to the injection point, which leads to the observed circular charge current. Because of the Rashba spin precession, the spin current oscillates as a function of the distance $r$ from the injection point, so that the charge current eventually changes direction as $r$ is increased. 
  While the charge current at long times (and fixed radius $r$) is almost entirely azimuthal, the current in the transient regime clearly has a significant radial component. This current occurs because of the different velocities for particle and hole excitations at finite spin voltage $V$. 
  Its magnitude depends approximately quadratically on $V$~\cite{SuppMat}, since it is affected by both the strength of the injected spin current and the average velocity difference. For realistic values of $V$, the radial current should thus be very small. It should also be noted that in a real system, the charge separation would be counteracted by the generated electrostatic potential, which is not accounted for in our model.

To make analytical predictions for the induced charge current that can be compared with the numerical results, it is more convenient to work with the continuous Rashba Hamiltonian
\begin{align}
  \hat{\mathcal{H}}_{R} &= \bm{\hat{p}}^2/2m + \alpha (\sigma^x \hat{p}_y - \sigma^y \hat{p}_x) \, .
\end{align}
By setting $m = 1/(2t_R)$ and $\alpha = -2 \lambda$, $\hat{\mathcal{H}}_R$ can be used to analyze the lattice version Eq.~\eqref{eqRashbaLattice} in the long-wavelength limit $k \rightarrow 0$. 
The continuum results are therefore applicable if the spin-orbit-coupling strength $\lambda $ is small and the Fermi energy $\varepsilon_F$ is close to the bottom of the electron bands (working at zero temperature, $\mu$ becomes the Fermi energy $\varepsilon_F$). 
In this regime, the wavenumber of the Rashba precession is $k_R = 2 \lambda $, which agrees with the widths of the observed current rings. 

\begin{figure}[t]
  \centering
  \includegraphics[width=0.99\linewidth]{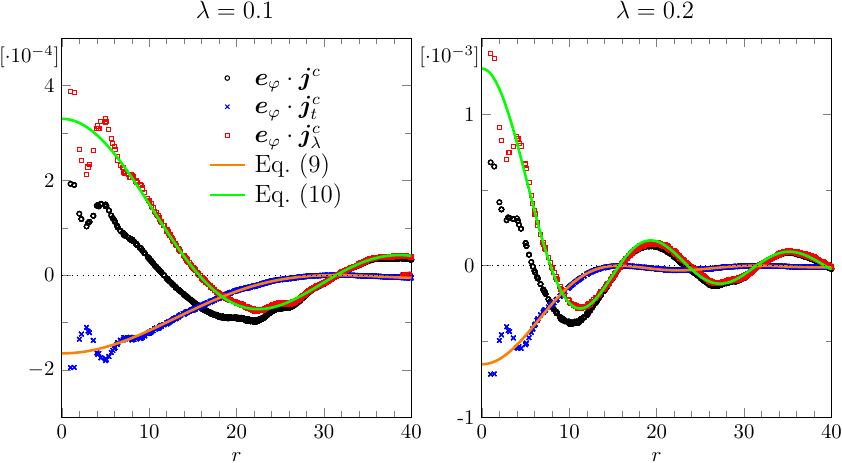} 
  \caption{Radial dependence of the azimuthal component of the charge current 
    at time $t = 45$. The solid lines are according to Eqs.~\eqref{eqsc1} and \eqref{eqsc2}. }
  \label{figR}
\end{figure}

Figure~\ref{figR} shows the radial dependence of the current for the largest simulated time $t = 45$ in more detail. 
Here, the charge current is separated into two parts, $\bm{\hat{j}}_t^c$ and $\bm{\hat{j}}_{\lambda}^c$, which are the terms proportional to $t_R$ and $\lambda$, respectively. 
Namely, we define
\begin{align}
  \bm{\hat{j}}_{t_R,\bm{r}}^c &= \frac{it_R}{2} \left[\bm{\hat{c}}_{\bm{r}}^\dagger (\bm{\hat{c}}_{\bm{r}-\bm{e}_x}^{\phantom{\dagger}} - \bm{\hat{c}}_{\bm{r}+\bm{e}_x}^{\phantom{\dagger}}) - \text{H.c.} \right] \bm{e}_x  \nonumber \\
  & \hspace*{4mm} + \frac{it_R}{2} \left[ \bm{\hat{c}}_{\bm{r}}^\dagger (\bm{\hat{c}}_{\bm{r}-\bm{e}_y}^{\phantom{\dagger}} - \bm{\hat{c}}_{\bm{r}+\bm{e}_y}^{\phantom{\dagger}}) - \text{H.c.} \right] \bm{e}_y \, , \label{eqjct} \\
    \bm{\hat{j}}_{\lambda, \bm{r}}^c &= \frac{\lambda}{2} \left[ \bm{\hat{c}}_{\bm{r}}^\dagger \sigma^y  (\bm{\hat{c}}_{\bm{r}+\bm{e}_x}^{\phantom{\dagger}} + \bm{\hat{c}}_{\bm{r}-\bm{e}_x}^{\phantom{\dagger}}) +\text{H.c.} \right] \bm{e}_x  \nonumber \\
  & \hspace*{4mm} - \frac{\lambda}{2} \left[ \bm{\hat{c}}_{\bm{r}}^\dagger \sigma^x (\bm{\hat{c}}_{\bm{r}+\bm{e}_y}^{\phantom{\dagger}} + \bm{\hat{c}}_{\bm{r}-\bm{e}_y}^{\phantom{\dagger}}) + \text{H.c.} \right] \bm{e}_y \, . \label{eqjclambda}
  \end{align}
The functional form of the two contributions can be explained using a semi-classical analysis in terms of wavepackets deflected by a spin-orbit force~\cite{SpinOrbitForce}. 
Let us consider the trajectory of an electron wavepacket at the Fermi energy $\varepsilon_F$ that has average momentum $\bm{p}$ and is initially centered at $\bm{r} = 0$ with the spin pointing up.
In addition to propagating in the direction of $\bm{p}$, it experiences an effective transverse force proportional to the $z$-component of the spin and the magnitude $p$ of the momentum.  
Since the spin oscillates with wavenumber $k_R$ because of the spin-orbit coupling, so does the deflecting force. 
This transverse movement corresponds to the spin-orbit part $\bm{j}_{\lambda}^c$ of the charge current. Furthermore, it causes the momentum $\bm{p}$ to no longer point in the radial direction  $\bm{e}_r = (x,y)^{\text{T}} / r$, so that the regular part $\bm{j}_t^c$ of the current obtains a finite component in the azimuthal direction $\bm{e}_\varphi = (-y,x)^{\text{T}} / r$ as well. 
By assuming that the injected spin current is composed in equal parts of wavepackets for spin-$\uparrow$ electrons and spin-$\downarrow$ holes that are evenly distributed over all directions, one obtains the following prediction for the charge current for long times $t$ and small $\lambda$: 
\begin{align}
  \bm{j}_t^c(r) &= j^z \frac{2 A}{k_R} \frac{\sin^2(k_R r /2 )}{r^2} \bm{e}_\varphi \, ,
  \label{eqsc1} \\
  \bm{j}_{\lambda}^c(r) &= -j^z A \frac{\sin(k_R r)}{r} \bm{e}_\varphi \, ,
  \label{eqsc2}
\end{align}
where $A=2 \lambda / (v_F \pi)$ is a constant that depends on the Fermi velocity $v_F = 2t_R \sqrt{4 + (\lambda / t_R)^2 + \varepsilon_F/t_R}$, and $j^z$ is the injected spin current. 
Inserting for $j^z$ the time-averaged value from the numerical simulations, 
we obtain excellent agreement with the numerically calculated charge current for $r \gtrsim 8$ (see Fig.~\ref{figR}), without any adjustable parameters. Deviations for small $r$ are likely due to the lattice discretization. 

Since the continuous Rashba Hamiltonian $\hat{\mathcal{H}}_{R}$ is symmetric under a simultaneous rotation of space and spin, the $z$-component of the total angular momentum
$\hat{J}^z=\hat{M} + \hat{S}^z$, where $\hat{M} = \hat{x} \hat{p}_y - \hat{y} \hat{p}_x$ is the orbital angular momentum, is conserved. 
While the lattice Hamiltonian $\hat{H}_R$ does not have this symmetry, we may expect the conservation of the total angular momentum to hold approximately, when the Fermi energy is small and the lattice model behaves similar to the continuum model. 
To be concrete, we define the orbital angular momentum on the lattice as
 $\hat{M} = \hat{x} \sin(\hat{p}_y) - \hat{y} \sin(\hat{p}_x)$. Using the first-quantized version of Eq.~\eqref{eqRashbaLattice}, $\hat{H}_R = -2 t_R [\cos(\hat{p}_x) + \cos(\hat{p}_y)] I - 2 \lambda [\sigma^x \sin(\hat{p}_y) - \sigma^y \sin(\hat{p}_x)]$, one then obtains from the Heisenberg equation: $d \hat{S}^z / d t  = -2 \lambda [\sin(\hat{p}_y) \sigma^y + \sin(\hat{p}_x) \sigma^x]$ and 
$d\hat{M} / dt = 2 \lambda [\cos(\hat{p}_x) \sin(\hat{p}_y) \sigma^y + \cos(\hat{p}_y) \sin(\hat{p}_x) \sigma^x]$. 
Obviously, $\hat{S}^z + \hat{M}$ is approximately conserved if we confine our analysis to states with small momenta $p$. To calculate $M$ in the interacting model numerically, we use the second-quantized expression 
\begin{align}
  \hat{M} &= - \frac{1}{2} \sum_{\bm{r}} \sum_{\mathclap{\sigma = \uparrow, \downarrow}} \left[ i x  \hat{c}_{\bm{r},\sigma}^\dagger \hat{c}_{\bm{r}+\bm{e}_y,\sigma}^{\phantom{\dagger}} - i y  \hat{c}_{\bm{r},\sigma}^\dagger \hat{c}_{\bm{r}+\bm{e}_x,\sigma}^{\phantom{\dagger}} + \text{H.c.} \right] \, .
\label{eqangmom}
\end{align}
Comparing with Eq.~\eqref{eqjct}, one can see that $\hat{M}$ is determined by the regular part $\bm{\hat{j}}_t^c$ of the charge-current-density operator $\bm{\hat{j}}^c$. 

\begin{figure}[t]
  \centering
  \includegraphics[width=0.99\linewidth]{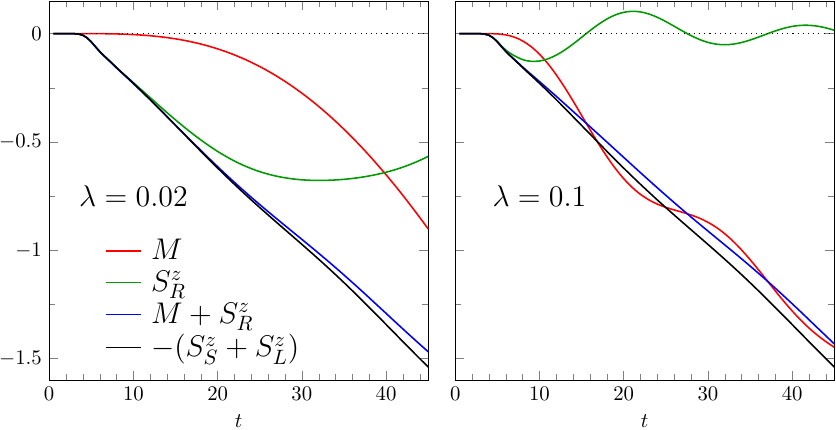}
  \caption{Time evolution of the $z$-component of the spin and orbital angular momentum in the Rashba system. 
 The total spins in the lead and the spin chain are denoted by $S_L^z$ and $S_S^z$, respectively. Therefore, the black line indicates the injected spin angular momentum. It would match the blue line if the total angular momentum was conserved. }
  \label{figM}
\end{figure}

When the spin current is injected, it increases the total angular momentum $J_R^z = S_R^z + M$ in the Rashba system. 
One might then ask how $J_R^z$ is composed of the spin $S_R^z$ and the orbital contribution $M$. 
Figure~\ref{figM} displays the numerical results for the time evolution of the angular-momentum expectation values.  
As noted above, the total angular momentum is not exactly conserved but the deviation is relatively small for $\varepsilon_F = -3.5$. 
  Initially, $M = S_R^z = 0$ because the spin current has not entered the Rashba system yet.  The delay before the angular momenta visibly change is in agreement with the expectation $N_S/v_S \approx 3.8$ based on the spinon velocity $v_S=J\pi/2$ in the infinite chain. 
For short times after the spin current has reached the Rashba system, $S_R^z$ makes up most of the angular momentum while $M$ remains approximately zero.
On longer timescales, however, $S_R^z$ can be seen to oscillate around zero, which means that eventually most of the injected spin angular momentum is converted to orbital angular momentum $M$.
With the same assumptions used to derive Eqs.~\eqref{eqsc1} and \eqref{eqsc2}, one obtains that both the amplitude and the period of the oscillations are proportional to the wavenumber $k_R$ of the Rashba precession. 
 The numerical results roughly agree with these predictions, except that the oscillations in $S_R^z$ and $M$ also appear to decrease with time.

We estimate the magnetic field generated by the current vortex following the Biot-Savart law of electromagnetism. By assuming $\lambda$, a lattice constant of $1\,$\AA, a hopping parameter $t_R=1\,$eV  and a linear dependence of the induced charge current on the spin voltage $V$, we obtain a field strength $B \approx V \cdot 10^{-5}\,$T/eV at the center.
For a realistic value of the spin voltage in the order of $10^{-4}\,$eV, this is about $10^{-9}\,$T and should therefore be within reach of experimental detection by scanning probe microscopy methods. 
To reach the necessary sensitivity, one could, e.g., use a nitrogen vacancy defect center in diamond as detector~\cite{PhysRevLett.106.080802}. 
We moreover expect that the magnetic field would be larger in a perhaps more realistic setup with a bundle of spin chains instead of a single chain. 
Finally, one could also consider injecting an ac spin current into the two-dimensional electron gas, in which case the current vortex would emit an electromagnetic field of similar strength. 

In conclusion, a charge-current vortex can be generated in a Rashba system by locally injecting a spin current. 
 The formation of the current vortex is accompanied by the conversion of the injected spin angular momentum to orbital angular momentum. 
 We demonstrated these effects for a generic model in which the spin current is transferred from an antiferromagnetic Heisenberg spin chain to a square-lattice Rashba system. 
In light of the recent realization of spin transport in the spin-chain material Sr$_2$CuO$_3$~\cite{SpinonSpinCurrent}, this model could be relevant from an experimental point of view. 
 Accurate time-dependent density-matrix renormalization-group results for the charge current were found to agree well with predictions from semi-classical considerations. 
The charge-current vortex induces an electromagnetic field, which may be observed experimentally.

 \textit{Acknowledgements} ---
 Density-matrix renormalization-group calculations were performed using the ITensor library~\cite{ITensor}. S. E. and F. L. are supported by Deutsche Forschungsgemeinschaft through project EJ 7/2-1 
 and FE 398/8-1, respectively.
J. F. is partially supported by the Priority Program of Chinese Academy of Sciences, Grant No. XDB28000000. 
S. Y and S. M. are supported by JST CREST (Grant No. JPMJCR19J4). S. Y. is financially supported by JSPS KAKENHI with Grant No. JP18H01183. S. M. is financially supported by JSPS KAKENHI (Grant No. JP20H01865) and JST CREST (Grant No. PMJCR1874).

\renewcommand{\thefigure}{S\arabic{figure}}
\renewcommand{\theequation}{S\arabic{equation}}
\setcounter{figure}{0}
\setcounter{equation}{0}

\renewcommand{\citenumfont}[1]{S#1}
\renewcommand{\bibnumfmt}[1]{[S#1]}

\onecolumngrid

\section{Supplemental Material}

\section{Block Lanczos transformation}

We use a Lanczos transformation that maps the two-dimensional Rashba system to a chain representation~\cite{SuppBLDMRG,SuppBLDMRGRashba}, which can be more efficiently numerically simulated with tensor-network techniques. 
The method amounts to applying the usual block Lanczos recursion to the matrix $H_R$ of the single-particle Rashba Hamiltonian on the lattice, using as initial vectors the spin-$\uparrow$ and spin-$\downarrow$ states of the site $\bm{r}_0$ coupled to the spin chain.
In second quantization, the transformed Rashba Hamiltonian is~\cite{SuppBLDMRGRashba}
\begin{align}
\hat{H}_R = &- \mu \sum_{j \geq 1} \sum_{\sigma = \uparrow,\downarrow} \hat{a}_{j,\sigma}^{\dagger} \hat{a}_{j, \sigma}^{\phantom{\dagger}} \nonumber + \sum_{j \geq 1} \sum_{\sigma = \uparrow,\downarrow} (\tilde{t}_j \hat{a}_{j, \sigma}^\dagger \hat{a}_{j+1, \sigma}^{\phantom{\dagger}} + \text{H.c.})  \, ,
\end{align}
where $\tilde{t}_j$ are bond-dependent real hopping parameters, $\hat{a}_{j,\sigma}$ are fermion annihilation operators and $\hat{a}_{1,\sigma} = \hat{c}_{\bm{r}_0,\sigma}$. Note that $\sigma$ is a pseudospin index that is equal to the physical spin only at the first site. 
The hopping parameters $\tilde{t}_j$ converge to $2t_R\sqrt{1+\lambda^2/(2t_R^2)}$ as $j \to \infty$. 

By applying the Lanczos transformation, the Hamiltonian becomes purely one-dimensional, consisting of a spin chain coupled to two semi-infinite tight-binding chains. 
For our numerical simulations, however, the system also needs to be truncated to a finite size. 
We do this by carrying out the Lanczos transformation for the first 150 sites and adding another 350 sites with the asymptotic hopping parameter as boundary conditions. Based on the calculated $\tilde{t}_j$, we estimate that the deviation from the exact hopping parameters is below $0.05\%$. 
The one-dimensional lead is truncated to 500 sites as well. 

The correlation functions in the chain representation of the Rashba system are used to calculate the expectation values in the original two-dimensional system by applying the reverse Lanczos transformation. Details on this procedure are given in Ref.~\cite{SuppBLDMRGRashba}. 

\section{Truncation error}
\begin{figure}[t]
\centering
\includegraphics[width=0.45\linewidth]{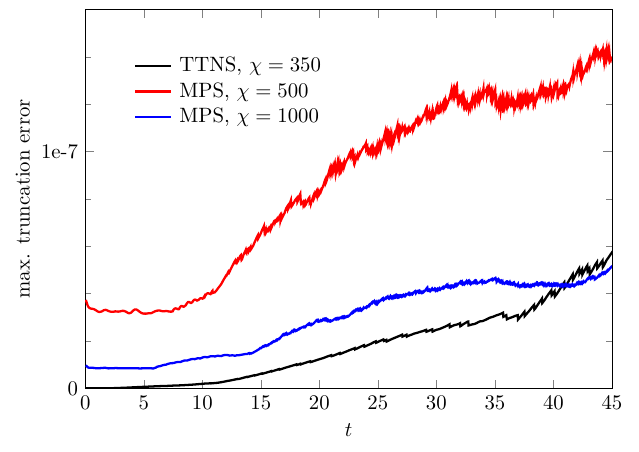}
\caption{
  Truncation error during the time-evolution with time step $\tau = 0.025$, using either a matrix-product state (MPS) or a tree-tensor-network state (TTNS) as ansatz. The bond dimension is denoted by $\chi$. Model parameters are $\lambda = 0.1$,  $N_S=12$, $J=t_L=2$, $\mu=-3.5$ and $V = 0.5$.
}
\label{fig_err}
\end{figure}

It is known that when a system is perturbed out of equilibrium, 
tensor-network descriptions gradually become less efficient, 
which is related to an increase in the entanglement entropy.  
Accordingly, there is a limit on the timescales that can be reached 
by our time-evolving block-decimation simulations. 
To make sure that the state of the system is still represented with sufficient accuracy, one can monitor the truncation error (defined as the sum of the discarded eigenvalues of the reduced density matrices) after the application of each Trotter gate. 
Figure~\ref{fig_err} displays the maximum truncation error for each time step. 
It remains below $10^{-7}$ for simulated times up to at least $t = 45$ if we use time steps of length $\tau = 0.025$ and a moderate maximum bond dimension $\chi = 350$ in the tensor network.  
The ansatz state, which is depicted in Fig.~2 of the main text, belongs to the class of tree-tensor network states (TTNS)~\cite{SuppTTNGeneral}. 
For comparison, we also show the truncation error for a matrix-product-state (MPS) ansatz, which requires much larger bond dimensions to reach a similar accuracy. 
It should be noted, however, that the reduction of the computational cost when using the tree-tensor network is somewhat
lessened by the worse scaling in the bond dimension for gates at the edges of the spin chain. 

\begin{figure}[t]
\centering
\includegraphics[width=0.5\textwidth]{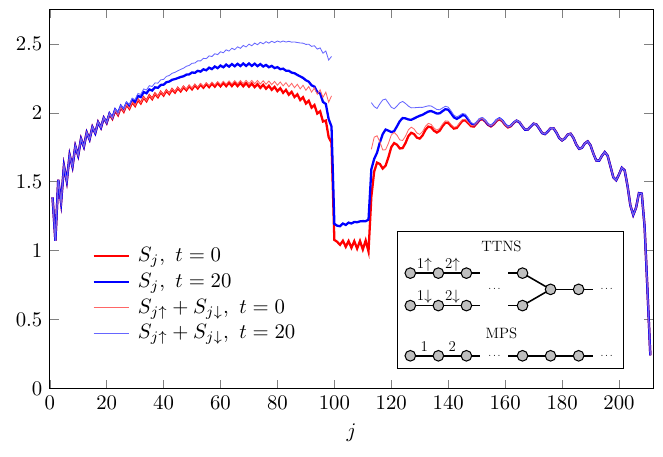}
\caption{
  Entanglement entropy for the bipartitions that appear in the matrix-product-state (MPS) and tree-tensor-network state (TTNS) representations. Model parameters are the same as in Fig.~\ref{fig_err} but the lengths of the tight-binding chains are reduced to $100$. The inset shows the definition of the bond indices, which are labeled starting from the one-dimensional lead, continuing with the spin chain and ending with the chain representation of the Rashba system. 
}
\label{fig_EE}
\end{figure}

In a loop-free tensor network, such as MPS or the more general TTNS, each bond $j$ corresponds to a partition of the system into two subsystems $A_j$ and $B_j$. The entanglement entropies $S_j = -\text{Tr}( \rho_{A_j}  \ln \rho_{A_j} ) = -\text{Tr}( \rho_{B_j}  \ln \rho_{B_j} )$, where $\rho_{A_j}$ and $\rho_{B_j}$ are the  reduced density matrices for the subsystems, 
roughly indicate how large the bond dimensions need to be to accurately represent a state. 
Figure~\ref{fig_EE} compares the entanglement entropies for MPS and TTNS in our transport simulations. For the TTNS, we use an additional spin-index to denote the bonds in each branch (see inset of Fig.~\ref{fig_EE}).
In terms of the entanglement entropies, the lower truncation error in the TTNS for similar bond dimensions may be explained as follows: The entanglement entropy $S_j$ is much higher in the leads than in the spin chain, so that the leads will require a larger average bond dimension in a MPS description. For the bonds in the individual lead branches of the TTNS, however, the entanglement entropy ($S_{j\uparrow}$ or $S_{j\downarrow}$) is significantly smaller. In fact, at time $\tau=0$, the entanglement entropy is almost halved, indicating that one approximately needs a bond dimension $\sqrt{\chi}$ instead of $\chi$. The leads, which make up most of the system, can therefore be simulated much more efficiently.

\section{\label{sec:continuity}Spin continuity equation}
In this section, we show the spin continuity equations for the Rashba model on a square lattice, and compare them against the numerical results presented in the main text. 
The Heisenberg equations of motion for the fermion operators $\bm{\hat{c}}_{\bm{r}}$ are 
\begin{align}
\frac{d}{d t} \bm{\hat{c}}_{\bm{r}}
	& = i t_R\left(
		  \bm{\hat{c}}^{}_{\bm{r}+\bm{e}_x}
		+ \bm{\hat{c}}^{}_{\bm{r}-\bm{e}_x}
		+ \bm{\hat{c}}^{}_{\bm{r}+\bm{e}_y}
		+ \bm{\hat{c}}^{}_{\bm{r}-\bm{e}_y}
	\right)
	- \lambda \left[
		\sigma^{y}  (\bm{\hat{c}}^{}_{\bm{r}+\bm{e}_x} - \bm{\hat{c}}^{}_{\bm{r}-\bm{e}_x}) 
		- \sigma^{x} (\bm{\hat{c}}^{}_{\bm{r}+\bm{e}_y} - \bm{\hat{c}}^{}_{\bm{r}-\bm{e}_y}) 
	\right]
. 
\end{align}
We ignored the chemical potential here, since it does not affect the time evolution of operators that conserve the total number of electrons. 
The time derivative of the total spin of a subset of sites $G$ is
\begin{align}
\frac{d}{d t} \hat{S}_G^{\alpha} = 
 \sum_{\bm{r} \in G} \frac{d}{d t} \hat{s}^{\alpha}_{\bm{r}}
	& = \sum_{\bm{r} \in G} \frac{1}{2} \left[
		  \bigg(\frac{d}{d t}\bm{\hat{c}}^{\dagger}_{\bm{r}} \bigg) \, \sigma^{\alpha} \bm{\hat{c}}^{}_{\bm{r}}
		+ \bm{\hat{c}}^{\dagger}_{\bm{r}} \sigma^{\alpha} \bigg( \frac{d}{d t} \bm{\hat{c}}^{}_{\bm{r}} \bigg)
	        \right] 
	 = \hat{\Gamma}_{t_R, x}^{\alpha}
		+ \hat{\Gamma}_{t_R, y}^{\alpha}
		+ \hat{\Gamma}_{\mathrm{soc}, x}^{\alpha}
		+ \hat{\Gamma}_{\mathrm{soc}, y}^{\alpha}
\label{eq:continuity}
,\end{align}
where we have introduced
\begin{align}
\hat{\Gamma}_{t_R, x}^{\alpha}
	& = -\frac{i t_R }{2} \sum_{\bm{r} \in G} \left[ \left(
		  \bm{\hat{c}}^{\dagger}_{\bm{r}+\bm{e}_x}
		+ \bm{\hat{c}}^{\dagger}_{\bm{r}-\bm{e}_x}
	\right) \sigma^{\alpha} \bm{\hat{c}}_{\bm{r}}
	- \bm{\hat{c}}^{\dagger}_{\bm{r}} \sigma^{\alpha} \left(
		  \bm{\hat{c}}^{}_{\bm{r}+\bm{e}_x}
		+ \bm{\hat{c}}^{}_{\bm{r}-\bm{e}_x}
	\right) \right]
, \\
\hat{\Gamma}_{t_R, y}^{\alpha}
	& = -\frac{i t_R}{2} \sum_{\bm{r} \in G} \left[ \left(
		  \bm{\hat{c}}^{\dagger}_{\bm{r}+\bm{e}_y}
		+ \bm{\hat{c}}^{\dagger}_{\bm{r}-\bm{e}_y}
	\right) \sigma^{\alpha} \bm{\hat{c}}_{\bm{r}}
	- \bm{\hat{c}}^{\dagger}_{\bm{r}} \sigma^{\alpha} \left(
		  \bm{\hat{c}}^{}_{\bm{r}+\bm{e}_y}
		+ \bm{\hat{c}}^{}_{\bm{r}-\bm{e}_y}
	\right) \right] ,
\end{align}
and
\begin{align}
\hat{\Gamma}_{\mathrm{soc}, x}^{\alpha}
	& = - \frac{\lambda}{2} \sum_{\bm{r} \in G} \left[ 
		(\bm{\hat{c}}^{\dagger}_{\bm{r}+\bm{e}_x} - \bm{\hat{c}}^{\dagger}_{\bm{r}-\bm{e}_x} ) \sigma^{y} \sigma^{\alpha} \bm{\hat{c}}_{\bm{r}}
		+ \bm{\hat{c}}^{\dagger}_{\bm{r}} \sigma^{\alpha} \sigma^{y} ( \bm{\hat{c}}^{}_{\bm{r}+\bm{e}_x} - \bm{\hat{c}}^{}_{\bm{r}-\bm{e}_x} )
	\right]
\label{eq:Gamma_soc_x}
, \\
\hat{\Gamma}_{\mathrm{soc}, y}^{\alpha}
	& = \frac{\lambda}{2} \sum_{\bm{r} \in G} \left[
		(\bm{\hat{c}}^{\dagger}_{\bm{r}+\bm{e}_y} - \bm{\hat{c}}^{\dagger}_{\bm{r}-\bm{e}_y} )\sigma^{x} \sigma^{\alpha} \bm{\hat{c}}_{\bm{r}}
		+ \bm{\hat{c}}^{\dagger}_{\bm{r}} \sigma^{\alpha} \sigma^{x} ( \bm{\hat{c}}^{}_{\bm{r}+\bm{e}_y} - \bm{\hat{c}}^{}_{\bm{r}-\bm{e}_y})
	\right] .
\label{eq:Gamma_soc_y}
\end{align}
In the continuum limit, $\hat{\Gamma}_{t_R,x}^{\alpha} + \hat{\Gamma}_{t_R, y}^{\alpha}$ corresponds to an integral of the divergence of the conventional spin current. 

\begin{figure}[tb]
\centering
\includegraphics[width=0.5\linewidth]{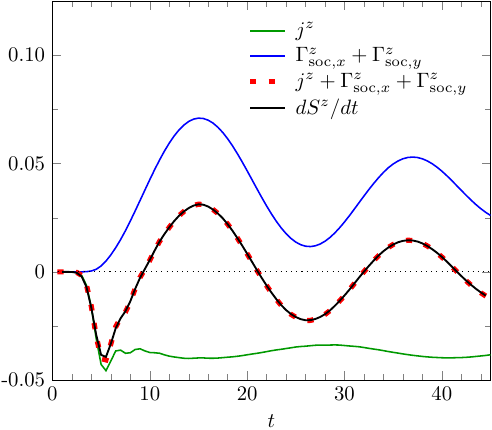}
\caption{\label{fig:S3} Numerical test of the spin continuity equation for $\alpha = z$ and $\lambda = 0.1$. 
The black line indicates the time derivative of the $z$-component of the total spin $\dot{S}^z$, the green line shows the injected spin current from the spin chain to the Rashba system, and the blue line denotes the rest contributions of the right hand side of the spin continuity equation. 
}
\end{figure}

Equation~\eqref{eq:continuity} can be used to check the accuracy of the numerical simulations. For this purpose, we evaluate the total spin $S^z$ (setting $G$ as the set of all sites) in the Rashba system at different times $t$, and then calculate the numerical derivative of $S^z(t)$. 
According to the continuity equation, we have $d \hat{S}^z /dt = \hat{j}^z + \hat{\Gamma}_{\text{soc},x}^z + \hat{\Gamma}_{\text{soc},y}^z$, where $\hat{j}^z$ is the spin current from the spin chain into the Rashba system 
(the terms $\hat{\Gamma}_{t_R,x}^z$ and $\hat{\Gamma}_{t_R,y}^z$ vanish for $\alpha = z$ when we sum over the whole system). 
As shown in Fig.~\ref{fig:S3}, this is in good agreement with the numerical results.

\section{Spin current polarized in-plane}

\begin{figure}[t]
\centering
\includegraphics[scale=0.49]{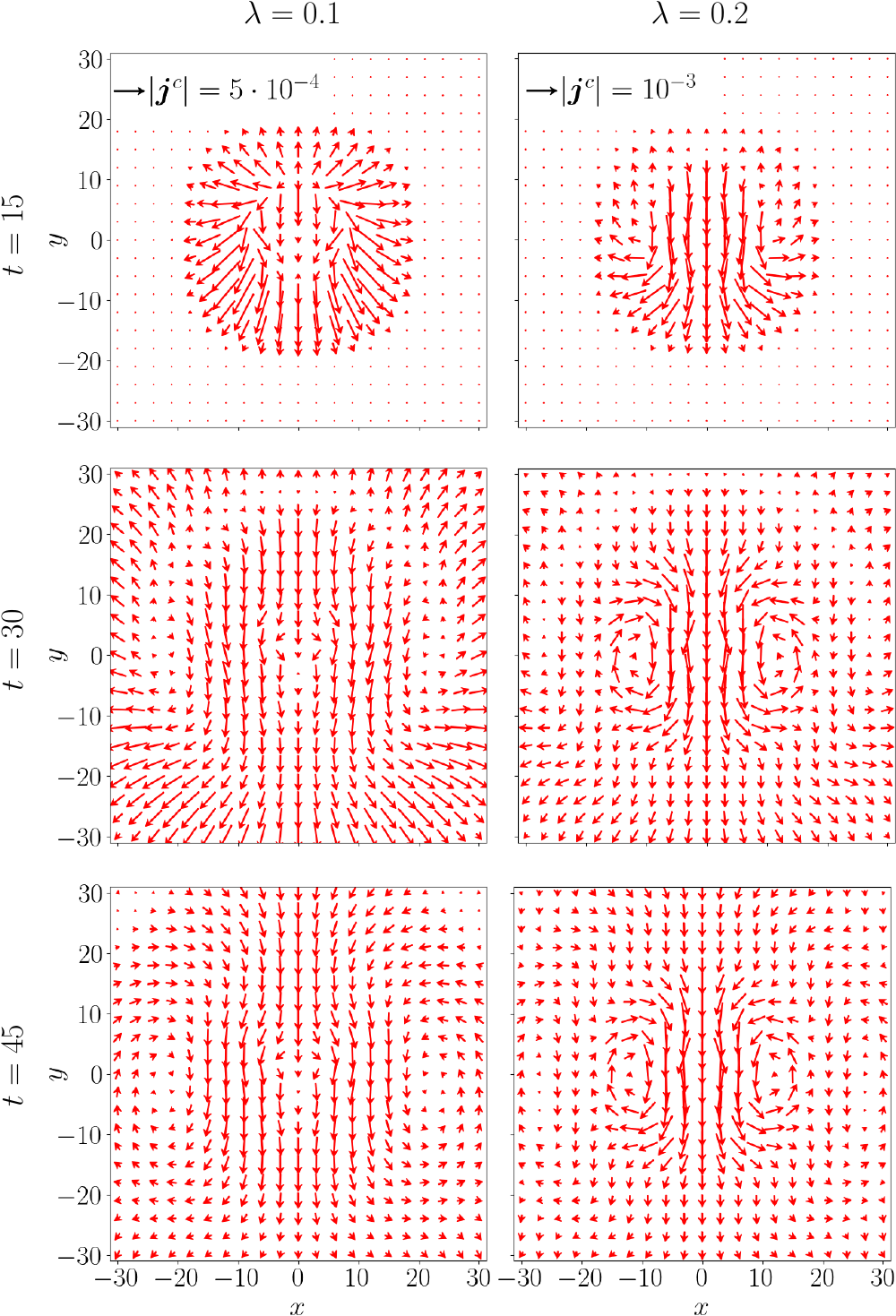}
\caption{\label{fig:S4} Charge current induced by an injected spin current polarized in the $x$-direction. For easier comparison, the other parameters are the same as in Fig.~3 of the main text, i.e., $N_S=12$, $J=t_L=2$, $\mu=-3.5$ and $V = 0.5$. }
\end{figure}

In the main text, we considered an injected spin current polarized in the $z$-direction, which is perpendicular to the plane of the Rashba system. Here, we present results for a spin current polarized in the $xy$-plane. 
From the effective one-dimensional model obtained by the Lanczos transformation it is clear that the magnitude of the spin current injected into the system is independent of the polarization direction~\cite{SuppBLDMRGRashba}. 
The induced charge current, on the other hand, is different. 
Figure~\ref{fig:S4} displays the charge current for $x$-polarization, 
with other parameters the same as in Fig.~3 of the main text. 
Instead of a vortex, there is now a net current in the negative $y$-direction. 
Since there is no rotational symmetry, it is more difficult to make precise predictions for the 
charge-current profile, but one can understand the main features  
by noting that (i) the inverse Rashba-Edelstein effect causes a charge current in the $y$-direction due to a distortion of the Fermi surface~\cite{SuppEdelstein,SuppMicroscopicEdelstein}, and (ii) the Rashba precession along the $x$-direction tilts the spin-current polarization out of the system plane, so that an inverse spin Hall current appears~\cite{SuppInverseSHE}. 
The latter effect leads to oscillations of the charge current along the $x$-axis. An important difference to the $z$-polarized spin injection is that the charge current is symmetric instead of antisymmetric under inversion about the $y$-axis. The orbital angular momentum is therefore zero. 
We also note that there is no conserved total angular momentum unless the spin current is polarized in the $z$-direction.

\section{Derivation of Eqs.~(9) and (10)}
The eigenfunctions and corresponding eigenvalues of the continuous Rashba Hamiltonian are 
\begin{align}
\psi_{\pm,\bm{p}}(\bm{r}) &= \frac{C}{\sqrt{2}} \begin{pmatrix} 1 \\ \pm i e^{i \varphi_{\bm{p}}} \end{pmatrix} e^{i \bm{p} \bm{r}} \, , \qquad \epsilon_{\pm, p} = p^2/2m \pm \alpha p \, ,
\end{align}
where $p=|\bm{p}|$, $\tan (\varphi_{\bm{p}}) = p_y/p_x$ and $C$ is a normalization constant. At fixed energy, the two bands are separated by a momentum difference $k_R=2 m \alpha$.

To describe the injection of a pure spin current, 
we consider pairs of spin-$\uparrow$ electron and spin-$\downarrow$ hole wavepackets that are initially localized at $\bm{r}=0$.
An electron wavepacket at the Fermi energy that is polarized in the $z$-direction and has average momentum $\bm{p}$ may be constructed as a superposition of states near $\psi_{+,\bm{p}(1-m\alpha/p)}$ and $\psi_{-,\bm{p}(1+m\alpha/p)}$, which have opposite spin orientation. As the wavepacket propagates, the momentum difference between the two contributions causes the spin to precess around an effective magnetic field proportional to $\bm{p} \times \bm{e}_z$. 
While the wavepacket of course spreads over time, we assume here that it can be treated as a classical particle with sharply defined momentum $\bm{p}$ and position $\bm{r}$.
Furthermore, we henceforth consider specifically a spin-$\uparrow$ electron wavepacket moving in the $x$-direction. 

Owing to the spin precession, the expectation value of the the $z$-component of the spin is $S^z(x) = \cos(k_R x)/2$. The spin-orbit interaction makes the electron experience an effective transverse force $-4 \alpha^2 m p S^z \bm{e}_y$~\cite{SuppSpinOrbitForce}, 
which deflects it in the $y$ direction but does not alter the momentum. 
Using $x = v_F t$, where $v_F = p/m$ is the Fermi velocity, one obtains   
$y(x) = -\frac{1}{2p}[1-\cos(k_R x)]$. 
This could also be derived by noting that the total angular momentum should be conserved, so that a change in $S^z$ has to be compensated by an opposite orbital angular momentum, i.e., $-y p  = 1/2 - S^z$. 
The azimuthal and radial components of the electron's velocity are
\begin{align}
  \bm{e}_\varphi \cdot \frac{d \bm{r}}{d t} &= \frac{1}{r} \left( \frac{1}{2m}[1 - \cos(k_R x)]  - \frac{k_R}{2 m}  x \sin(k_R x) \right) \nonumber \\
  & \approx \frac{1}{mr} \sin^2(k_R r /2) - \frac{k_R}{ 2 m} \sin(k_R r) 
  \label{eqapprox1}
\end{align}
and
\begin{align}
  \bm{e}_r \cdot \frac{d \bm{r}}{d t} &= \frac{1}{r} \left(x v_F + \frac{k_R}{4 p m }[1 - \cos(k_R x)] \sin(k_R x)   \right) \nonumber \\
& \approx v_F \, ,
  \label{eqapprox2}
\end{align}
respectively. In the second lines of Eqs.~\eqref{eqapprox1} and \eqref{eqapprox2}, we assumed that $y/x \ll 1$, which is valid for a small spin-orbit force, or long-enough distances $r$. 
For a spin-$\downarrow$ hole, the radial component is the same while the azimuthal one is reversed.
As described in the following, the above equations 
can be used to estimate the charge current induced by the spin injection. 

In our approximation, the spin current $j^z$ into the Rashba system is equal to the rate with which the pairs of spin-$\uparrow$ electron and spin-$\downarrow$ hole wavepackets are injected.
Since the wavepackets are evenly distributed over all directions and move radially approximately with $v_F$, their density is $j^z/(2 \pi r v_F)$. 
The charge-current density, taking both spin-$\uparrow$ electrons and spin-$\downarrow$ holes into account, is then 
\begin{align}
  \bm{j}^c(r) &= \frac{j^z}{\pi v_F r} \; \left[ \frac{1}{mr} \sin^2(k_R r /2) - \frac{k_R}{ 2 m} \sin(k_R r) \right] \bm{e}_\varphi \, .
  \label{eqjcpred}
\end{align}
There is no radial component, since the electron and hole contributions cancel each other.
The first term in Eq.~\eqref{eqjcpred} can be identified as the regular part of the charge current, and the second term as the additional spin-orbit part. 
By setting $m = 1/(2t_R)$ and $\alpha = -2\lambda$, we obtain Eqs.~(9) and (10) of the main text.

\section{Spin injection by spin pumping}
So far, we have studied a setup, in which a spin current is injected locally by coupling the Rashba system to a quantum spin chain. 
It turned out that the charge current in the Rashba system could be understood in a simple semi-classical picture that does not directly reference the spin chain. A natural question is then, if the charge-current vortex also appears
in a simpler model without spin chain, which may be easier to handle numerically. 
In this section, we demonstrate that, indeed, a current vortex is also induced, when a $z$-polarized spin current is injected by coupling the site at $\bm{r}_0$ to a precessing classical spin. 
The corresponding Hamiltonian is $\hat{H}(t) = \hat{H}_R + \hat{H}_C(t)$, where $\hat{H}_R$ is the same Rashba Hamiltonian as in Eq.~(1) of the main text and
\begin{equation}
\hat{H}_C(t) = \frac{J}{2} \sum_{\nu = x,y,z} S_\nu(t) ( \bm{\hat{c}}_{\bm{r}_0}^{\dagger} \sigma^\nu  \bm{\hat{c}}_{\bm{r}_0}^{\phantom{\dagger}} ) \, ,
\end{equation}
with
\begin{equation}
  S_x = S \cos(\omega t \Theta(t)) , \quad S_y = -S \sin(\omega t \Theta(t)) , \quad S_z = 0 \, ,
\end{equation}
is the coupling to a spin with magnitude $S$ that points initially in the $x$-direction and at time $t = 0$ starts precessing around the $z$-axis with angular frequency $\omega$. Here, we set $S=1$ for simplicity. The precession of the spin induces a spin current with polarization direction $\bm{S} \times \dot{\bm{S}} = -\bm{e}_z$ like in the spin-chain model with $V>0$. An advantage of this Hamiltonian is that it is quadratic and thus can be efficiently simulated without tensor-network techniques.

\begin{figure}[t]
\centering
\includegraphics[width=0.8\textwidth]{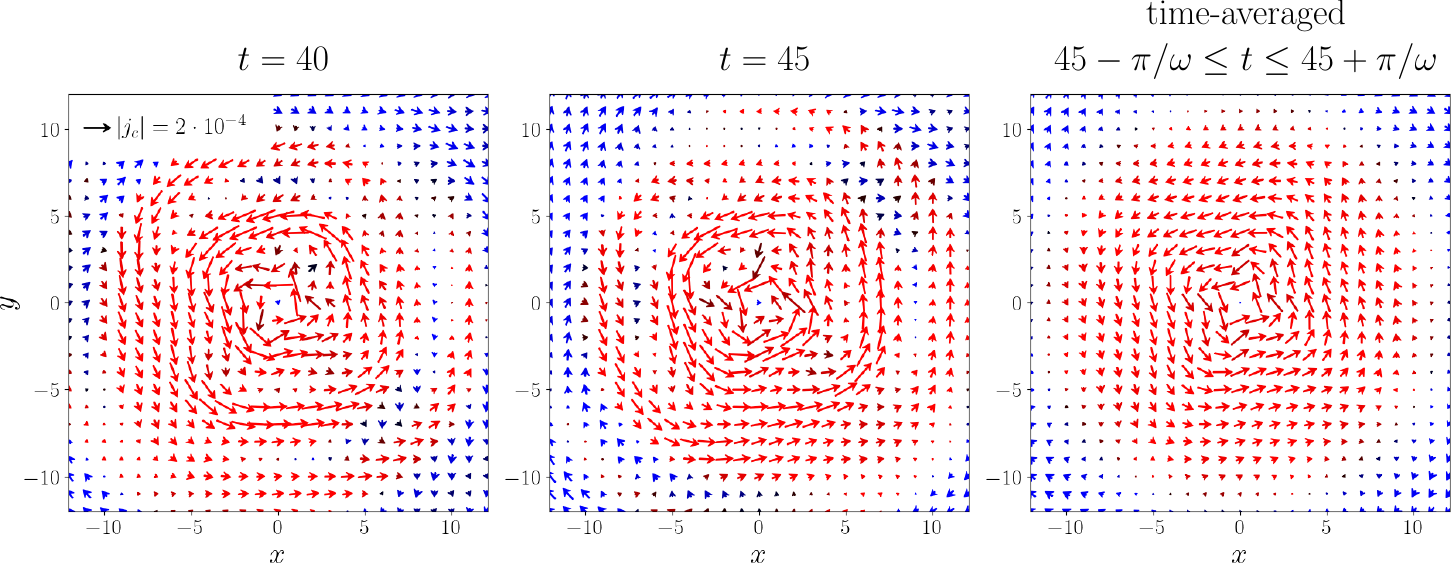}
\caption{Charge-current profile for the simplified model with classical spin. The parameters are $\lambda=0.1$, $J=4$ and $\omega =0.25$. Note that a smaller region is shown than in Fig. 3 of the main paper.}
\label{fig_jct}
\end{figure}

As before, we apply a Lanczos transformation to the system to obtain a one-dimensional representation. The Hamiltonian can then be further simplified by using a gauge transformation $\hat{a}_{j \uparrow} \rightarrow e^{i\omega t \Theta(t)} \hat{a}_{j \uparrow}$ that
turns the Hamiltonian into
\begin{align}
\hat{H}(t) &= \hat{H}_R + \hat{H}_C(t < 0) + \Theta(t) \omega \sum_{j} \hat{a}_{j \uparrow}^{\dagger} \hat{a}_{j \uparrow}^{\phantom{\dagger}} \, .
\end{align}
By setting $\omega = V/2$, the induced spin current has approximately the same magnitude as in the model with the spin chain, provided that $J$ is chosen appropriately. 
We diagonalize the single-particle Hamiltonian for $t < 0$ and $t > 0$ to calculate the correlation functions in the one-dimensional representation, which are then used to obtain the current densities in the original two-dimensional system. 

A difference to the junction model in the main text is that the Hamiltonian does not have a rotational symmetry around the $z$-axis. If we average the expectation values over a period $2 \pi / \omega$ of the oscillation, however, we obtain a very similar charge-current profile. 
The time-dependent deviations come from the correlation functions that are mixed in the pseudo-spin basis. These correlation functions converge with time, 
except for a periodic phase factor from the gauge transformation, which leads to the observed cancellation when averaged over time. 
Figure~\ref{fig_jct} demonstrates these fluctuations for $\lambda = 0.1$ and other parameters the same as in the main text.

\section{Displaced charge}
\begin{figure}[b]
\centering
\includegraphics[width=0.65\textwidth]{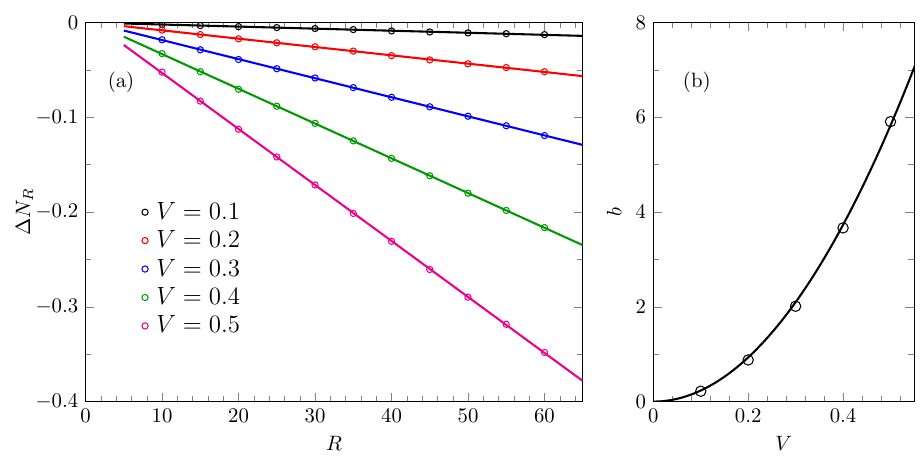}
\caption{(a) $\Delta N_R$ for the model with classical spin and parameters $\lambda = 0.1$, $\mu=-3.5$ and $J=4$. Lines are linear fits. (b) Fit parameter $b$ in $\Delta N_R(R) = a - b R$ ($a,b\in \mathbb{R}$) as a function of $V$. The line is a quadratic fit $(b =  \text{const} \cdot V^2)$.}
\label{fig_displN}
\end{figure}
In this section, we analyze the displaced charge due to the transient radial charge current. We use the model from the previous section and consider the change $\Delta N_R$ of the charge inside a circle with radius $R$ around the injection point:
\begin{align}
\Delta N_R &= \lim_{t \to \infty} \sum_{r \leq R} \left[\langle \hat{c}_{\vec{r}}^{\dagger}\hat{c}_{\vec{r}}^{\phantom{\dagger}} \rangle (t) - \langle \hat{c}_{\vec{r}}^{\dagger}\hat{c}_{\vec{r}}^{\phantom{\dagger}} \rangle (0) \right] \, .
\end{align}
Figure~\ref{fig_displN} shows $\Delta N_R$ for different radii $R$ and spin voltages $V$. 
The displaced charge appears to depend linearly on $R$, with a slope that increases quadratically with $V$. 
Qualitatively, the results can be understood as follows: 
Particle and hole excitations induced near the classical spin are propagating with sligthly different velocities because of the finite $V$ and the nonlinear electron dispersion, leading to different densities of the excitations, i.e., more (less) particles for large (small) $r$. 
The spin voltage $V$ affects both the strength of the injected spin current and the difference in the velocities of the excitations, which leads to the approximately quadratic dependence of the displaced charge on $V$.

\end{document}